\documentclass[structabstract]{aa}  
\usepackage{graphicx}
\usepackage{amsmath,amssymb}
\usepackage{txfonts}
\usepackage{natbib}
\usepackage{color}
\usepackage[colorlinks=true,citecolor=blue]{hyperref}
\usepackage{xspace}
\begin{document}
   \title{Calibration of quasi-static aberrations in exoplanet direct-imaging instruments with a Zernike phase-mask sensor}

   \author{M. N'Diaye\inst{\ref{inst:LAM}, \ref{inst:STScI}, \ref{inst:PHASE}}
     \and
     K. Dohlen\inst{\ref{inst:LAM},\ref{inst:PHASE}}
     \and
     T. Fusco\inst{\ref{inst:ONERA},\ref{inst:LAM},\ref{inst:PHASE}}
     \and
     B. Paul\inst{\ref{inst:ONERA},\ref{inst:LAM},\ref{inst:PHASE}}
   }

   \institute{Aix Marseille Universit\'e, CNRS, LAM (Laboratoire d'Astrophysique de Marseille) UMR 7326, 13388, Marseille, France\\ 
   \email{[mamadou.ndiaye,kjetil.dohlen]@oamp.fr}\label{inst:LAM}
         \and
         Office National d'Etudes et de Recherches A\'erospatiales, D\'epartement d'optique th\'eorique et appliqu\'ee, 29 avenue de la division Leclerc, 92322 Ch\^atillon, France\label{inst:ONERA}
         \and
         Space Telescope Science Institute, 3700 San Martin Drive, Baltimore MD 21218, USA\label{inst:STScI}
         \and
	Groupement d'Int\'er\^et Scientifique PHASE (Partenariat Haute R\'esolution Angulaire Sol-Espace)\label{inst:PHASE}
           }

   \date{Received ********; accepted ********}

% \abstract{}{}{}{}{} 
% 5 {} token are mandatory
 
  \abstract
   {Several exoplanet direct-imaging instruments (VLT-SPHERE, Gemini Planet Imager, etc.) will soon be in operation, providing original data for comparative exoplanetary science to the community. To this end, exoplanet imagers use an extreme adaptive optics (XAO) system to correct the atmospheric turbulence and provide a highly corrected beam to a near-infrared (NIR) coronagraph for suppression of diffracted stellar light. The performance of the coronagraph is, however, limited by the non-common path aberrations (NCPA) due to the differential wavefront errors existing between the visible XAO sensing path and the NIR science path and leading to residual speckles that hide the faintest exoplanets in the coronagraphic image.}
   {Accurate calibration of the NCPA in exoplanet imagers is mandatory to correct the residual, quasi-static speckles remaining in the coronagraphic images after XAO correction in order to allow the observation of exoplanets that are at least $10^6$ fainter than their host star. Several approaches have been developed during these past few years to reach this goal. We propose an approach based on the Zernike phase-contrast method operating in the same wavelength as the coronagraph for the measurements of the NCPA between the optical path seen by the visible XAO wavefront sensor and that seen by the NIR coronagraph.}
   {This approach uses a focal plane phase mask of size $\sim \lambda/D$, where $\lambda$ and $D$ denote the wavelength and the telescope aperture diameter, respectively, to measure the quasi-static aberrations in the upstream pupil plane by encoding them into intensity variations in the downstream pupil image. The principle of this approach as described in several classical optical textbooks is simplified by the omission of the spatial variability of the amplitude diffracted by the phase mask. We develop a more rigorous formalism, leading to highly accurate measurement of the NCPA, in a quasi-linear way during the observation.}
   {With prospects of achieving subnanometric measurement accuracy with this approach for a static phase map of standard deviation 44\,nm rms at $\lambda=1.625\,\mu$m (0.026\,$\lambda$), we estimate a possible reduction of the NCPA due to chromatic differential optics by a factor ranging from 3 to 10 in the presence of adaptive optics (AO) residuals compared with the expected performance of a typical current-generation system. This would allow a reduction of the level of quasi-static speckles in the detected images by a factor 10 to 100, thus correspondingly improving the capacity to observe exoplanets.}
   {}


   \titlerunning{Zernike phase mask sensor}
   \maketitle

%
%________________________________________________________________

%%%%%%%%%%%%%%%%%%%%%%%%%%%%%%%%%%%%%%%%%%%%%%%%%%%%%%%%%%%%%
\section{Introduction}\label{sec:intro}
Following the recent images of the exoplanets HR8799 b, c, d \citep{2008Sci...322.1348M,2011ApJ...729..128C,2011ApJ...741...55S,2011ApJ...739L..41G} and $\beta$ Pic b \citep{2009A&A...506..927L,2010Sci...329...57L,2011ApJ...736L..33C,2011A&A...528L..15B}, the astronomical community has high expectations for future discoveries and studies in comparative exoplanetology. In the next few years, high-contrast imaging and spectroscopy will provide many clues to the frequency, diversity, and habitability of exoplanets, the formation and evolution of planetary systems, the relationship between brown dwarfs and planets, etc. \citep{2009ARA&A..47..253O}. The forthcoming instruments, VLT-SPHERE \citep{2008SPIE.7014E..41B}, Gemini Planet Imager \citep[GPI, ][]{2008SPIE.7015E..31M}, Subaru-SCExAO \citep{2010SPIE.7736E..71G}, and Palomar P1640 \citep{2011PASP..123...74H}, will start lifting the veil with the direct imaging of young or massive gaseous planets that are 10$^{6}$ fainter than their host star at a few tenths of an arcsecond. Space missions will also be concerned since the future James Webb Space Telescope \citep{2008SPIE.7010E..19C} will form the images of planets of about two Jupiter masses around M-type stars, offering about 10$^5$ contrast level at half an arcsecond from a host star \citep{2010ASPC..430..167C}. The next decade will see the emergence of new exoplanet imagers, such as EPICS \citep{2008SPIE.7015E..46K}, for the future European Extremely Large Telescope \citep[E-ELT,][]{2008SPIE.7012E..43G} on the ground and coronagraphic telescopes in space \citep{2010SPIE.7731E..68G,2010SPIE.7731E..67T,2012ExA...tmp...11B}, for the study and spectral analysis of extrasolar planets from 10$^7$ to 10$^{10}$ times fainter than their host stars at a few hundredths of an arcsecond.

However, the huge contrast ratio at a small angular separation (from $10^5$ to $10^{10}$ in the visible and near-infrared at less than one arcsecond) between a host star and its planetary companion make the observation of such an object very challenging \citep{2010exop.book..111T}. Implementation of several techniques, including extreme adaptive optics \citep[XAO, ][]{2006OExpr..14.7515F,2008SPIE.7015E..31M,2010SPIE.7736E..71G,2011PASP..123...74H}, stellar coronagraphy \citep[e.g., see the review of concepts in ][]{2006ApJS..167...81G}, and post-processing methods \citep[e.g.][]{2000PASP..112...91M,2002ApJ...578..543S,2006ApJ...641..556M,2007ApJ...660..770L,2008OExpr..1618406M,2008A&A...489.1345V}, is required to disentangle the photons of the planetary companions from those of its host star. An XAO system will form images at the resolution limit for a ground-based telescope with high Strehl ratio (better than 90\%), compensating for the effects of the atmospheric turbulence. A coronagraph will strongly suppress the stellar signal, removing the telescope diffraction effects. Post-processing methods will attenuate the speckles present in the coronagraphic image due to ripples in the residual wavefront after adaptive optics correction to make further improvements, necessary to reduce these residual speckles which are quasi-static wavefront deformations due to non-common path aberrations (NCPA) between the optical path seen by the visible wavefront sensor and that seen by the near-infrared (NIR) coronagraph. For instance with SPHERE, a pointing error smaller than 0.5\,mas rms was specified to ensure a satisfying centering of the star image on the coronagraph and to reach the scientific objectives of the instrument \citep{2006SPIE.6269E..24D}. In addition, \citet{2006IAUS..232..149S} estimates that wavefront errors (wfe) lower than $\lambda/280$ and $\lambda/2800$ with $\lambda$ denoting the wavelength of study, or equivalently 6.0\,nm rms and 0.6\,nm rms in the H-band ($\lambda=1.65\,\mu$m), will lead to $10^7$ and $10^9$ contrast thresholds, respectively.

The current-generation instruments, SPHERE and GPI, will respectively rely on the phase-diversity method \citep{2007JOSAA..24.2334S,2008OExpr..1618406M} and an interferometric approach \citep{2010SPIE.7736E.179W} to measure the non-common path aberrations. They are part of the many methods proposed in the past few years, while other concepts rely on interferometry \citep{1994Natur.368..203A,2008A&A...489.1389N,2008A&A...488L...9G,2010A&A...509A..31G}, speckle-nulling technique \citep{2007Natur.446..771T}, coronagraphic phase diversity \citep{Sauvage:10}, analysis of the light blocked by a coronagraph \citep{2009ApJ...693...75G,2011PASP..123.1434V}, or the Zernike phase contrast principle \citep{2003SPIE.5169..309B, 2011SPIE.8126E..11W, 2012SPIE.8450E..0NN}. Several post-processing methods have also been developed to exploit the chromatic behavior of speckles \citep{2000PASP..112...91M,2002ApJ...578..543S}, using their rotation with respect to the sky \citep{2006ApJ...641..556M} or based on algorithms of optimal recombination of images \citep{2007ApJ...660..770L}.
 
We here consider the Zernike sensor based on the phase-contrast method \citep{1934MNRAS..94..377Z}. The interest of pursuing the study of this concept is illustrated well by \citet{2005ApJ...629..592G}, who claims that it is "ideal" since it has a photon noise sensitivity factor ($\beta_P$ in his notation) of unity where competing concepts, such as the fixed pyramid sensor, have a sensitivity factor of $1/\sqrt{2}$, and the Shack-Hartmann sensor, with the assumptions used in the paper, reaches a minimum factor of 3. We have analyzed the case of using this ideal sensor as a slow wavefront sensor applied to the measurement and real-time calibration of the chromatic component of the NCPA between the optical path seen by the visible XAO wavefront sensor and the one seen by the NIR coronagraph. Our approach is to use the Zernike wavefront sensor in the NIR to minimize these chromatic differential aberrations.

\begin{figure}[!ht]
\centering
\resizebox{\hsize}{!}{
\includegraphics{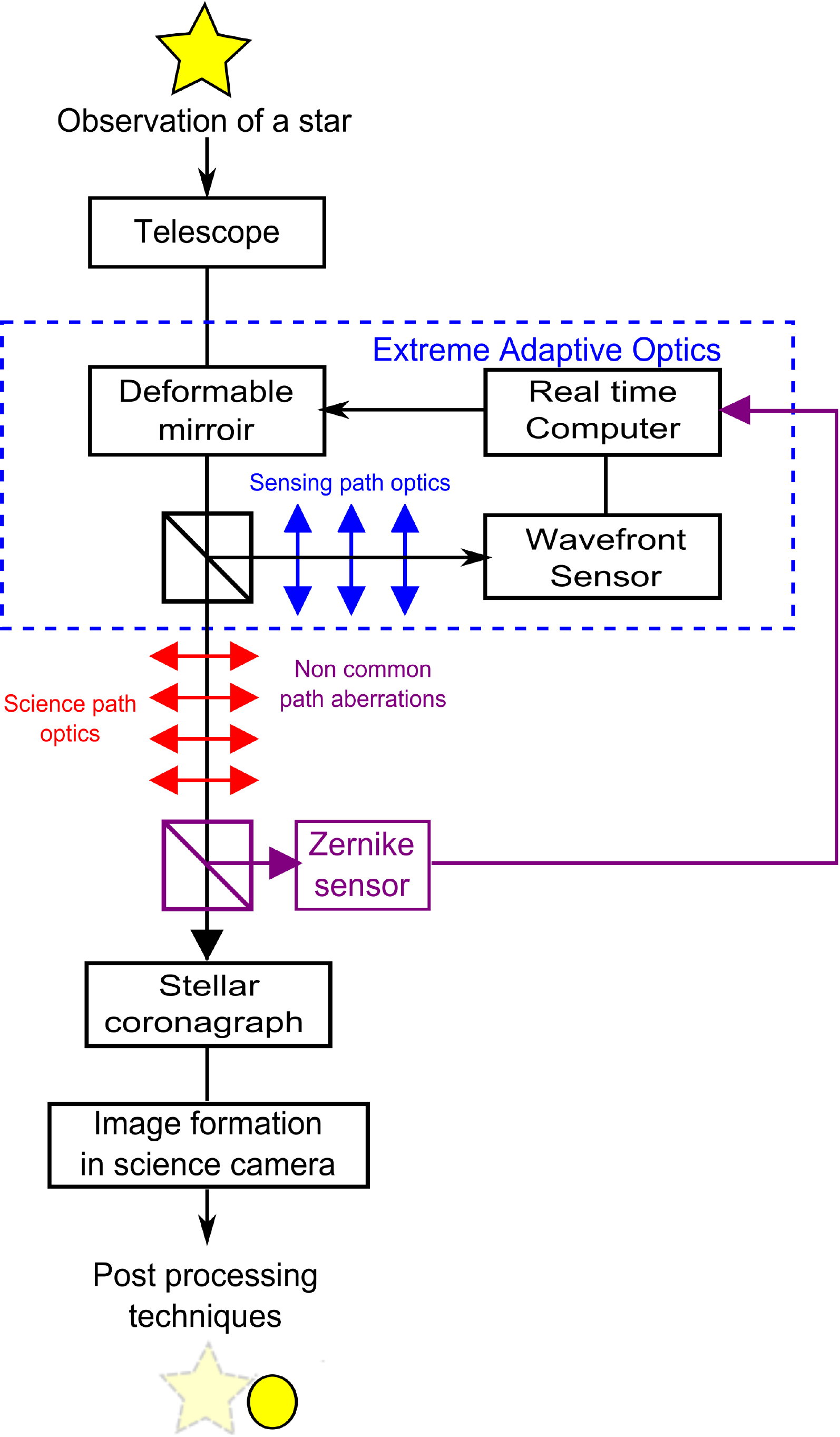}
}
\caption{General diagram of exoplanet direct imaging including a Zernike sensor. By deploying a NIR sensor, it may be placed closer to the coronagraph entrance than the high-order and visible XAO sensor, and thus not suffer from NCPA due to optics located between the two beam splitters}
\label{fig:HCIT_scheme}
\end{figure}

By taxing a small part of the science beam (<10\%) as close as possible upstream of the coronagraph (see Fig. \ref{fig:HCIT_scheme}) and by using long exposures (>1s), we benefit from the excellent noise-propagation properties of this sensor to correct these quasi static aberrations in parallel with science observations. Clearly, the use of a beam splitter before the coronagraph implies that the Zernike sensor does not see exactly the same aberration function as the coronagraph. However, since the differential optics are limited to a single, fixed component whose optical quality can be optimized by the use of high-quality optics, the differential aberrations can be very small (typically 1-2\,nm rms) and stable in time. This point is particularly important and represents the main advantage over the classical approach, used notably in SPHERE, where the chromatic component of the NCPA is calibrated at intervals of several hours using phase diversity. Since the chromatic differential optics (represented as double arrows in Fig. \ref{fig:HCIT_scheme}) between the visible XAO wavefront sensor and the NIR coronagraph consists of physically long optical paths (several meters) containing fixed, as well as variable, optical elements (e.g., atmospheric dispersion correctors using rotating prism elements), their contributions to NCPA are necessarily time variable, leaving non-negligible, chromatic differential aberrations in the coronagraph plane. In the proposed Zernike sensor, we will still rely upon classical calibration methods such as phase diversity (new developments in coronagraphic phase diversity [\citealt{2012SPIE.8446E..8BP}] can be of particular interest in this context), but the lifetime of this calibration is expected to be much longer, probably extending well beyond one night. 

%_____________________________________________________________
%_____________________________________________________________

The GPI instrument approaches this problem by implementing a real-time, post-coronagraphic wavefront sensing system (HOWFS) based on a Mach-Zehnder interferometer setup. Again, a beam splitter taxes a fraction of the science light, but since the separation occurs after the coronagraph, it does not introduce differential aberrations. Potentially, this system therefore does not require external calibration. The cost of such a system lies in its complexity, however, the Mach-Zehnder interferometric system is very alignment sensitive, and a piezo-electric phase modulation system is included, etc. In comparison, the Zernike sensor, as proposed for SPHERE, would fit in a cylindrical envelope roughly 300\,mm long and 50\,mm in diameter (including the beam splitter but excluding the detector cryostat), containing no moving elements. 
 
\citet{2003SPIE.5169..309B} propose a practical implementation of the Zernike sensor for real-time AO operation in combination with a Shack-Hartmann wavefront sensor for boot-strapping, and \citet{2011SPIE.8126E..11W} propose a phase-shifting version using an original piston system. \citet{2012SPIE.8450E..0NN} presented the first results of a development of the sensor based on fixed, ion-beam machined phase masks, arguing the obtention of high-quality masks with small ($\lambda/D$) dimensions. A similar concept has been tried earlier for the measurement of phasing errors in segmented telescopes \citep{2006SPIE.6267E.102D} and demonstrated on the sky with the VLT \citep{2010ApOpt..49.4052S,2011ApOpt..50.2708V}, using a mask with a size roughly equal to the atmospheric seeing-disk diameter. It is interesting to note that the Zernike sensor can be seen as a compact equivalent of the Mach-Zehnder interferometer with an amplitude or phase mask inserted in the interfering arms \citep{1994Natur.368..203A,2002ESOC...58..113L,2004EAS....12...33D}.
It is also noteworthy that the Zernike sensor can be used with any aperture shape and pupil obscuration (shadow of the secondary mirror and spiders, gaps between segments, etc.) and that it is sensitive to phasing errors in segmented telescope systems. It therefore constitutes a promising wavefront-sensing option for future space and ground-based instruments like EPICS, receiving light through segmented primary collectors such as the European Extremely Large Telescope. On a shorter timescale, its insertion into current high-contrast imagers like SPHERE on the VLT will allow online coronagraph-plane wavefront sensing as a complement to the currently implemented phase diversity method at a minimal cost in terms of system modifications.

This paper addresses the performance of the Zernike sensor in the context of a SPHERE upgrade as described above in order to move forward practical implementation and on-sky demonstration of the concept. A recollection of the mathematical formalism is followed by an analysis of the impact of various error sources, such as chromatic effects and AO residuals, leading to constructing a complete error budget.  While the principles of the Zernike phase-contrast method are explained in classical textbooks \citep{hecht1987optics,1992ost..book.....M,1996ifo..book.....G,1999po..book.....B}, the mathematical description used there is not fully complete. By introducing our notation we provide a comprehensive treatment, considering also the case of deviations from the classical $\pi$/2 phase shift, which can be particularly useful for measuring very small aberrations.

To estimate the precision obtained with this method, we have identified error sources, including detection noise, AO residuals, and chromatic effects, and built an error budget as a function of target flux. This detailed study of measurement errors constitutes a major contribution to the knowledge of the Zernike sensor.

We have excluded investigation of online measurements of variable amplitude aberrations since these are not considered limiting for current-generation systems (SPHERE, GPI, etc.) and since these instruments do not have means for correcting such aberrations (i.e., a second deformable mirror). However, given a slight increase in complexity, allowing observation of a pupil image not seen through the Zernike mask simultaneously with the Zernike pupil image, online amplitude measurements will be possible.

In Sect. \ref{sec:ZSPM}, we describe the formalism of the Zernike phase-mask sensor, underlining its simplicity, sensitivity, and quasi-linearity for a direct reconstruction of the phase map related to the measurement of the intensity in the pupil. Numerical simulations illustrate the reconstruction of a static phase map with this approach in Sect. \ref{sec:reconstruction}, while phase-map reconstruction in the presence of AO residuals is investigated in Sect. \ref{sec:AOresidual}. Chromaticity effects, such as finite spectral bandwidth or the difference between design and measurement wavelengths, are analyzed in Sect. \ref{sec:chromaticity}. We finally discuss the application of the Zernike sensor to an exoplanet direct imaging instrument in Sect. \ref{sec:real_system}, proposing an error budget in order to estimate the ultimate performance limits of the system. 
 
%%%%%%%%%%%%%%%%%%%%%%%%%%%%%%%%%%%%%%%%%%%%%%%%%%
\section{The Zernike phase mask sensor}\label{sec:ZSPM}  % \label{} allows reference to this section
\subsection{Principle}\label{subsec:principle}
\begin{figure}[!ht]
\centering
\resizebox{\hsize}{!}{
\includegraphics{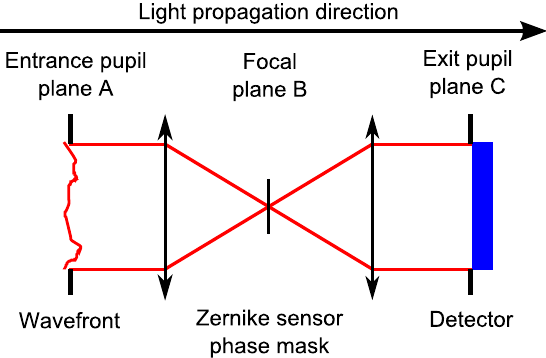}
}
\caption{Layout of the Zernike sensor for an accurate reconstitution of the residual phase in the framework of a high-contrast exoplanet imager.} 
\label{fig:ZSPM_scheme}
\end{figure}

The scheme of the Zernike phase mask concept is given in Fig. \ref{fig:ZSPM_scheme}. Residual wavefront errors after the XAO system are contained in the complex amplitude of the electric field at the entrance pupil plane A of the Zernike sensor. A phase mask in the form of a small circular depression in a glass plate is placed at the following focal plane B where the star image is formed, introducing a phase change for the complex amplitude of the central part of the star image going through the mask. This leads to interference between the electric fields going through and outside the phase disk in the relayed pupil in plane C, producing an intensity pattern that is related to the wavefront aberrations. The exact intensity encoding of wavefront errors depends on the size and depth of the mask, and quasi-linearity is achieved with an adequate choice of these parameters.

We note that this interferometric phenomenon is closely related to the nulling observed in the phase mask stellar coronagraph \citep{1997PASP..109..815R} where a circular $\pi$-phase disk is used to remove diffracted starlight. The objective of the Zernike sensor, however, is different since it aims at a coding of the residual aberrations as pupil plane intensity variations rather than optimal coronagraphic extinction.    

%_____________________________________________________________
%_____________________________________________________________

\subsection{Formalism}\label{subsec:formalism}
The classical textbook description of the mathematical formalism of the Zernike sensor \citep{hecht1987optics,1992ost..book.....M,1996ifo..book.....G,1999po..book.....B} as reported in \citet{2011SPIE.8126E..11W} is often somewhat simplified, neglecting the variability of the reference wave's complex amplitude across the pupil.

\begin{figure}[!ht]
\centering
\resizebox{\hsize}{!}{
\includegraphics{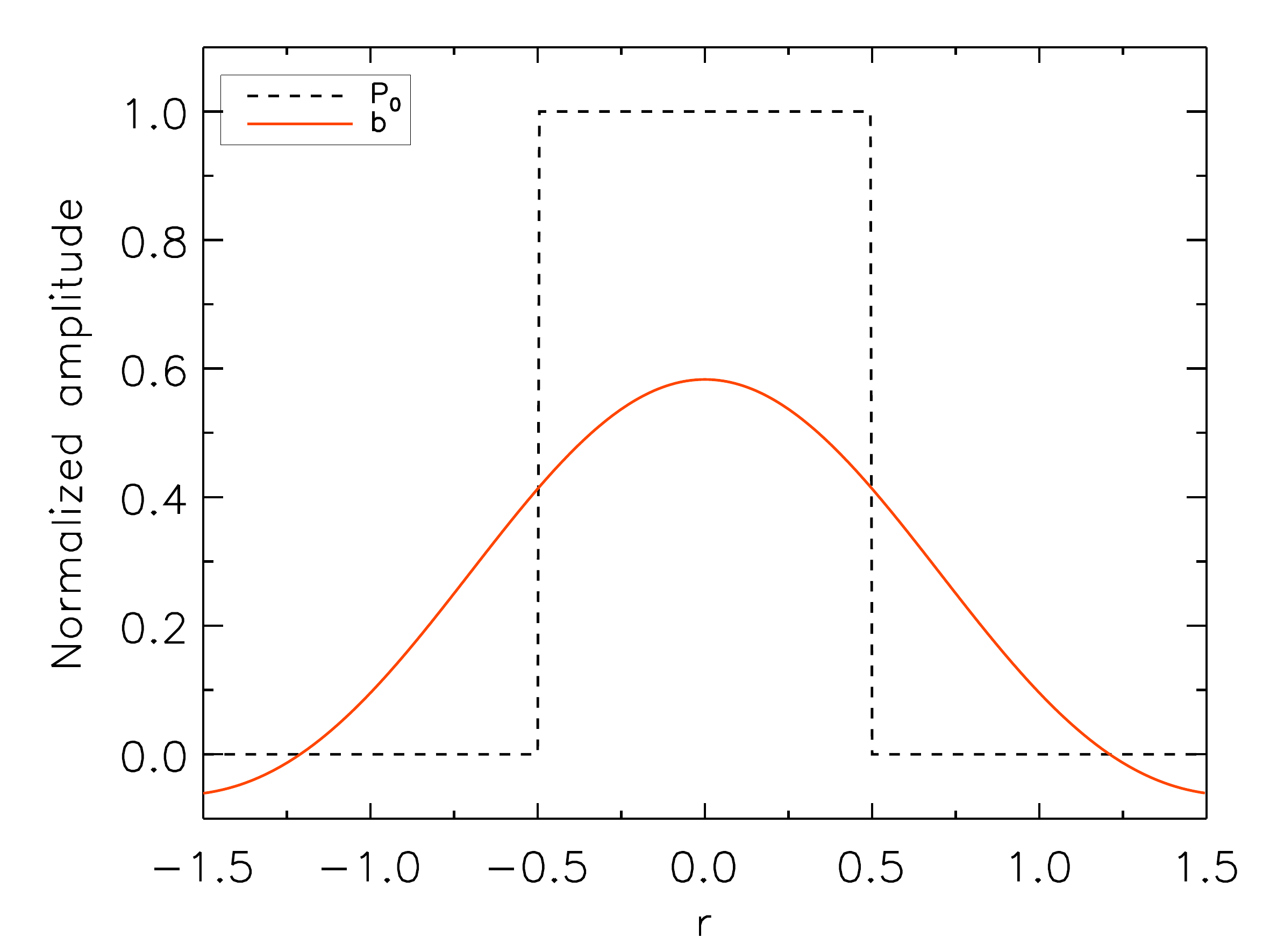}
}
\caption{Radial profile of the amplitude $b$ diffracted by a mask of size 1.06\,$\lambda/D$ and phase shift $\theta=\pi/2$. The dashed line defines the entrance pupil function $P_0$.} 
\label{fig:b_plot}
\end{figure}

In the following, we describe the formalism of the Zernike sensor. For the sake of clarity, we omit the position vectors $\textbf{r}$ and $\boldsymbol{\rho}$ in the pupil and focal planes, their modulus $r$ and $\rho$, and the wavelength $\lambda$, and $\mathcal{F}$ symbolizes the Fourier transform operator in which we include the Fourier optics scaling factor $1/\lambda f$, with $f$ the telescope focal length \citep{1996ifo..book.....G}. The Fourier transform is also shorthand written with a hat: $\mathcal{F}[A]=\widehat{A}$. The operator $\otimes$ denotes the convolution product.\\
The complex amplitude of the electric field $\Psi_A$ at the aperture (plane A) is given by
\begin{equation}
\Psi_A=Pe^{i \varphi} = P_0\,(1-\epsilon)\,e^{i \varphi}\,,%
\label{eq:ampltude_A}
\end{equation}
in which the real functions $P$ and $\varphi$ describe the amplitude and phase, respectively, of the entrance pupil. The phase function is assumed to have a zero mean, and $P$ is normalized such that $P_0$ defines the telescope aperture shape, equal to 1 inside the pupil and 0 elsewhere and $\epsilon$ is a zero-mean amplitude error function. 
The phase mask is located in the following focal plane, denoted plane B. Its amplitude transmission function $t$ can be written as
\begin{equation}
t=1 - (1 - e^{i\theta})\,M,
\label{eq:transmission_ZSPM}
\end{equation}
in which $M$ defines the top-hat function of the phase mask equal to 1 for $|\rho|<d/2$ and 0 otherwise, $d$ denoting the mask diameter. The term $\theta$ represents the phase shift introduced by the mask.\\
The complex amplitude of the field $\Psi_B$ after the mask is then given by
\begin{equation}
\Psi_B=t \,\widehat{\Psi}_A\,,
\label{eq:amplitude_B}
\end{equation}
and so the complex amplitude in the exit pupil following the mask is
\begin{equation}
\begin{split}
\Psi_C & = \widehat{\Psi}_B\\ 
& = \Psi_A - (1-e^{i\theta})\,\widehat{M} \otimes \Psi_A\,.
\end{split}
\label{eq:ampltude_C}
\end{equation}
For masks $M$ smaller than the Airy disk, $\widehat{M}$ becomes a smooth function that is broader than the pupil. In highly corrected systems, and since $\epsilon$ and $\varphi$ have zero mean, the convolution product can be approximated to
\begin{equation}
b = \widehat{M} \otimes \Psi_A  
\simeq \sqrt{\mathcal{S}} \widehat{M} \otimes P_0 =  \sqrt{\mathcal{S}} b_0\,,
\end{equation}
where $\mathcal{S}=(1-\sigma^2/2)^2$ is the Strehl ratio [\citealp{marechal1947}] and $\sigma^2$ the wavefront variance. This is a real function, smoothly varying across and beyond the pupil, see Fig. \ref{fig:b_plot}. Knowing both pupil geometry and mask geometry, $b_0$ can be computed once and for all. With typical mask diameters close to $\lambda/D$, $b$ has a profile similar to that of an Airy pattern twice the size of the pupil, see Fig. \ref{fig:b_plot}. The presence of $\mathcal{S}$ in this expression is problematic for the absolute measurement of finite wfe, but in the vicinity of zero aberration, which is the regime we seek through closed-loop correction, it will be close enough to unity to be ignored.
We note that the spatial variability of $b$ is ignored both by classical authors \citep{1992ost..book.....M,1999po..book.....B} and by contemporary authors \citep{2011SPIE.8126E..11W}, treating this as a constant. Maintaining the spatial distribution of $b$ is important in our analysis since ignoring it will introduce spurious aberrations.

%_____________________________________________________________
%_____________________________________________________________

We can now rewrite the exit pupil electric field as
\begin{equation}
\begin{split}
\Psi_C & = \Psi_A - (1-e^{i\theta})\,b\\
& = a + b e^{i\theta}\,,
\end{split}
\label{eq:ampltude_C_ab}
\end{equation}
with $a=\Psi_A-b$.
The electric field in pupil plane C is seen to represent the interference of the wavefront $a$ emanating from outside of the image plane mask with $b$, a smooth reference wavefront, emanating from within the mask. As in classical interferometry we therefore expect, as noted and exploited by Zernike, gaining him a Nobel prize, that phase modulations in plane A show up as intensity variations in plane C.

\begin{figure}[!ht]
\centering
\resizebox{\hsize}{!}{
\includegraphics{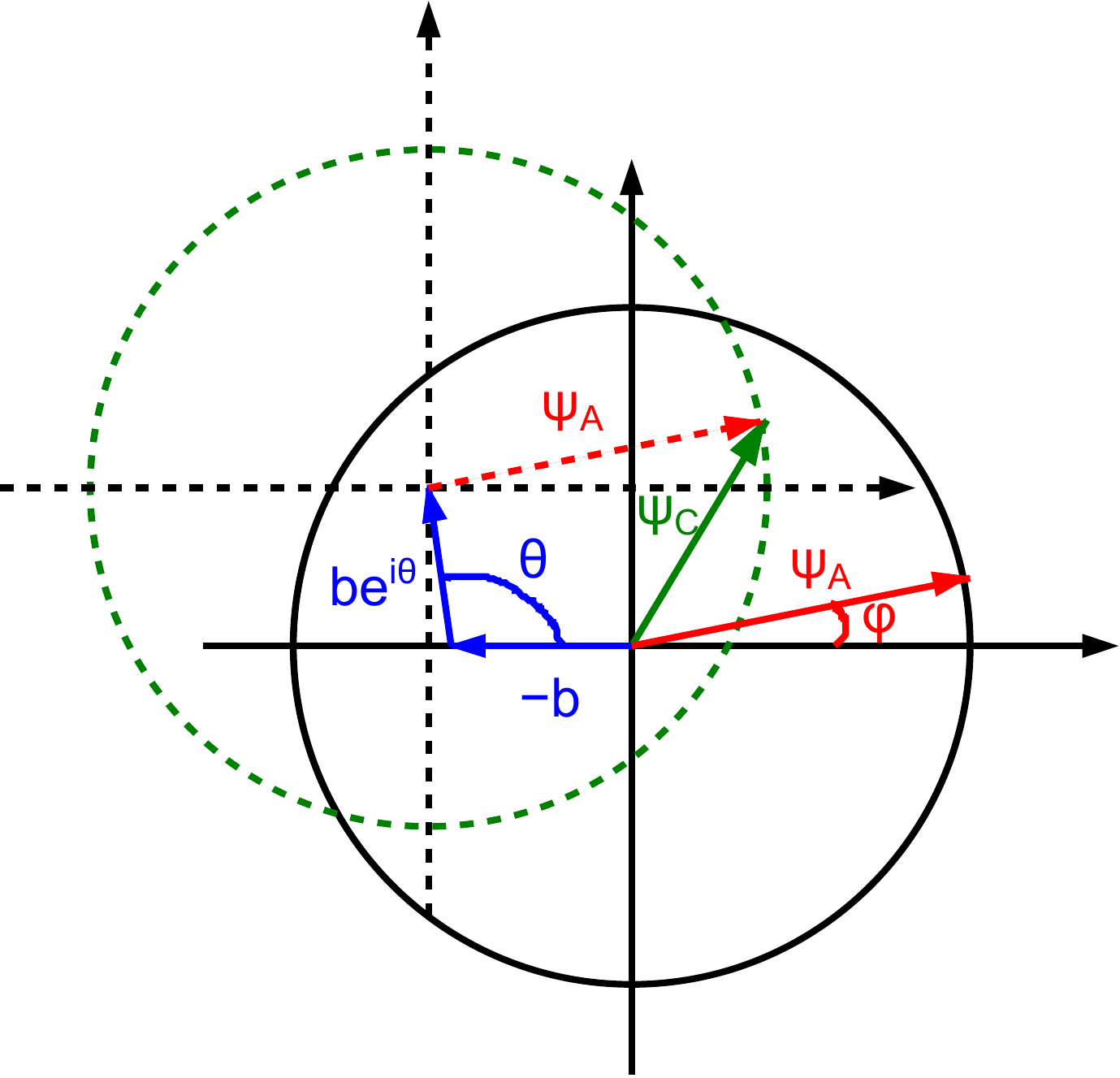}
}
\caption{Schematic representation in Argand diagram of the different parameters involved in Eq. (\ref{eq:ampltude_C_ab}) to describe the exit pupil plane amplitude $\Psi_C$.} 
\label{fig:Argand_diagram}
\end{figure}

Corresponding to phase error $\varphi$, Eq. (\ref{eq:ampltude_C_ab}) can be represented in the Argand diagram as a shifted circle. The unit circle centered on the origin representing $\Psi_A$ is shifted to a new center at $-b\,(1-e^{i\theta})$, see Fig. \ref{fig:Argand_diagram}. The intensity in the exit pupil $I_C$ is equal to the square of the length of the vector $\Psi_C$ joining the corresponding point on the shifted circle to the origin: 
\begin{equation}
I_C = |\Psi_C|^2\,.
\label{eq:intensity_C}
\end{equation}

%_____________________________________________________________
%_____________________________________________________________

\begin{figure}[!ht]
\centering
\resizebox{\hsize}{!}{
\includegraphics{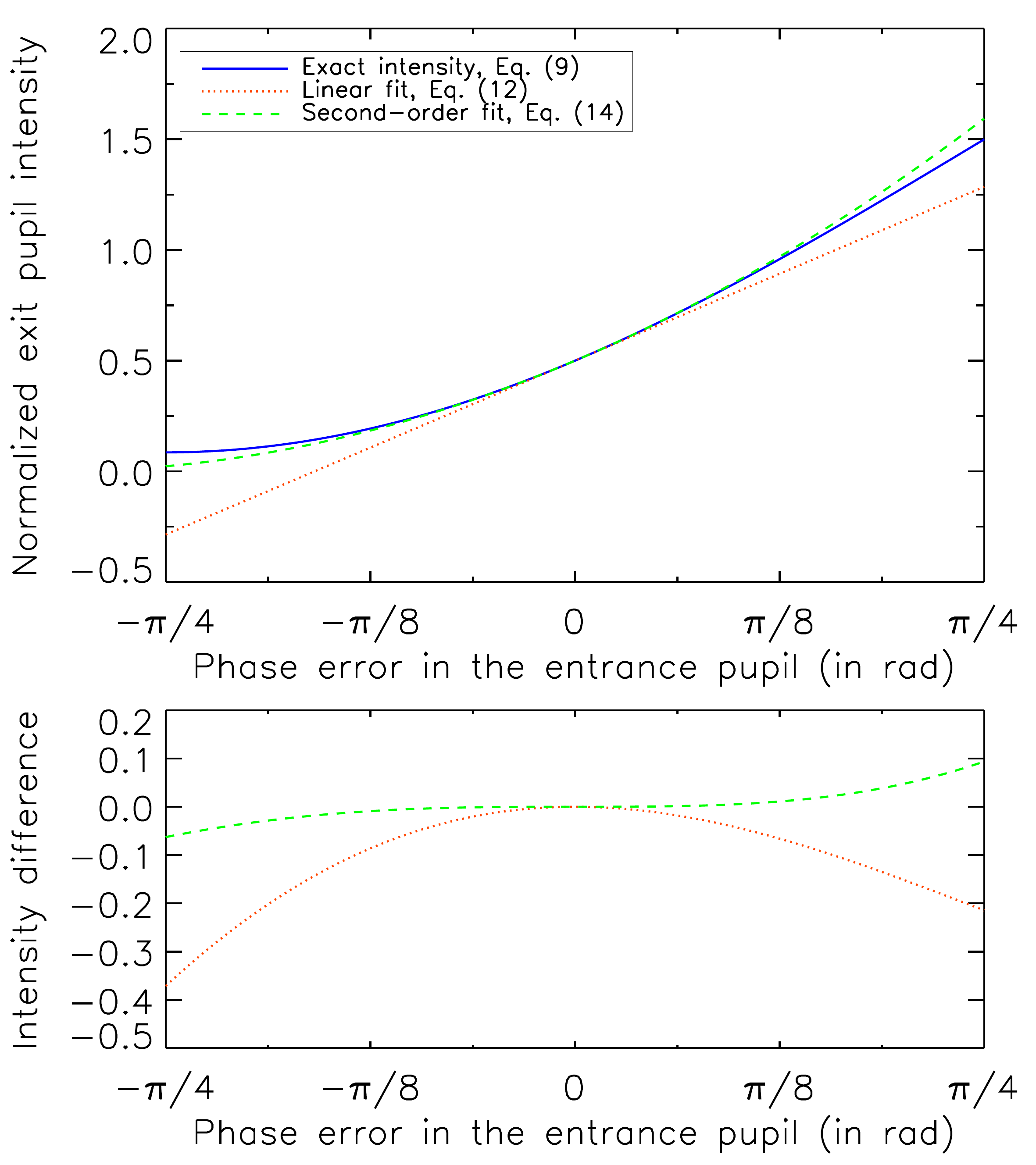}
}
\caption{Profile of the exit pupil intensity $I_c$ as a function of the phase error in the entrance pupil $\varphi$ for the case $P=1$, $b=0.5$, and $\theta=\pi/2$. \textbf{Top}: Comparison of the exact relationship between phase and intensity (blue) with linear (red) and second-order (green) approximations. \textbf{Bottom}: Error of the linear (red) and second-order (green) approximations.} 
\label{fig:comparison_approximations}
\end{figure}

As expected, the length of this vector is indeed related in a quasi-linear way to the phase errors in the entrance pupil for a certain range of $\varphi$. Figure \ref{fig:comparison_approximations} plots the exit pupil intensity as a function of phase error for a typical mask geometry.

%_____________________________________________________________
%_____________________________________________________________

By accounting for amplitude errors in the pupil, the scheme described above becomes somewhat more cluttered, but noting that if we remove the mask from plane B (actually just shifting it a distance equal to a few times its diameter), we obtain an exit pupil intensity of $I_{C0} = |\Psi_A|^2 = P^2 = P_0^2\,(1-\epsilon)^2$, from which the pupil amplitude function P, hence $\epsilon$, can be extracted during the calibration stage. 

Inverting Eq. (\ref{eq:intensity_C}) yields $\varphi$ as a function of $I_C$. Expressing the complex exponentials as cosines and sines and using Eqs. (\ref{eq:ampltude_A}) and (\ref{eq:ampltude_C_ab}), we find that
\begin{equation}
\begin{split}
I_C & = \left [P\cos\varphi - b(1-\cos\theta)\right ]^2 + \left [P\sin\varphi + b \sin\theta \right ]^2\\
& = P^2 + 2b^2(1-\cos\theta) + 2Pb\left [\sin\varphi\sin\theta - \cos\varphi(1-\cos\theta) \right ]\,.
\end{split}
\label{eq:intensity_C_develop}
\end{equation}
For very small phase errors, we can consider a linear case where we only maintain first-order terms of $\varphi$ in the Taylor expansion. Then we find
\begin{equation}
I_C = P^2 + 2b^2(1-\cos\theta) + 2Pb\left [\varphi\sin\theta - (1-\cos\theta)\right ]\,.
\label{eq:intensity_C_expand_lin}
\end{equation}
Then, the phase can be recovered from the measured image as
\begin{equation}
\varphi = \frac{1}{\sin\theta}\left [\frac{I_C}{2Pb} - \frac{P}{2b} + \left (1 - \frac{b}{P} \right )(1-\cos\theta)\right ]\,.
\label{eq:phi_solu_lin}
\end{equation}
For the case of $b=0.5$ and $\theta=\pi/2$, including small amplitude errors deduced from an unfiltered pupil image, this reduces to
\begin{equation}
\varphi = I_C - 0.5  - \epsilon\,.
\label{eq:phi_solu_lin_simp}
\end{equation}
A more accurate expression is obtained by maintaining the second order of $\varphi$ in the Taylor expansion. Then $I_C$ can be written as
\begin{equation}
I_C = P^2 + 2b^2(1-\cos\theta) + 2Pb\left [\varphi\sin\theta - (1-\varphi^2/2)(1-\cos\theta)\right ]\,.
\label{eq:intensity_C_expand}
\end{equation}

Solving this second-order equation is tedious and not very rewarding since it does not simplify well. For the purpose of illustration we therefore provide here the solution corresponding to the classical case where $\theta=\pi/2$ and ignoring amplitude errors: $P=1$ within the pupil. Then,
\begin{equation}
I_C = 1 + 2b^2 + 2b\,(\varphi^2/2 + \varphi - 1)\,,
\label{eq:intensity_C_simp}
\end{equation}
for which we get the solution
\begin{equation}
\varphi = -1 + \sqrt{3-2b-(1-I_C)/b}\,.
\label{eq:phi_solu}
\end{equation}
For $b=0.5$, the phase error is simply given by $\varphi=-1+\sqrt{2I_C}$.

Figure \ref{fig:comparison_approximations} compares the exact expression (Eq. (\ref{eq:intensity_C_develop}) in blue solid line) with this simplified expression (Eq. (\ref{eq:phi_solu}), second-order in green dashed line) and its linear version (Eq. (\ref{eq:phi_solu_lin_simp}), first-order in red dotted line) in the case where $P=1$, $b=0.5$, and $\theta=\pi/2$. Clearly, the second-order approximation represents the exact expression in the $\pm \pi/4$ range well. The linear expression is significantly less accurate, but could be interesting in a closed-loop, null-error context.

A value of $b$ close to 0.5 over the pupil is achieved using a phase mask with angular diameter of 1.06\,$\lambda/D$ in the case of a circular aperture, see Fig. \ref{fig:b_plot}. This recalls the case of a phase-mask coronagraph, where a mask of this dimension but with a $\pi$ phase shift achieves perfect nulling in the center of the coronagraphic exit pupil \citep{1997PASP..109..815R}.

Our study is limited to circular apertures, but the formalism of the Zernike sensor is valid for any aperture shape and obscuration (shadow of the secondary mirror and spiders, gaps between segments, etc.) since the intensity measurement is made inside the geometric pupil. With the Zernike phase mask, like the Roddier \& Roddier phase mask, the light diffracted by an obscuration in the entrance pupil and modified by the mask interference effects will remain within the re-imaged obscurations in the relayed pupil, making the geometric pupil free of light contamination, unaltered for the exit pupil intensity measurements and wavefront reconstitution inside it. Phase discontinuities caused by segment phasing errors are also perfectly coded by this approach, making the Zernike sensor a very promising system for the NCPA and piston measurements in EPICS, the future exoplanet imager for the segmented E-ELT \citep{2008SPIE.7015E..46K}.  

\subsection{Noise propagation}\label{subsec:noise_propagation}
To investigate the ultimate performance of this concept, we consider error propagation due to measurement noise and other effects in the case of operation near zero phase error. In this case we are safely within the linear regime described by Eq. (\ref{eq:phi_solu_lin_simp}). Since the intensity used in this expression is normalized by the average entrance pupil flux, $I_C = S/\overline{S_0}$, where $S$ is the measured signal in the Zernike exit pupil and $\overline{S_0}$ is the average signal in the entrance pupil (both in number of photo-electrons per pixel):
\begin{equation}
	\varphi = S/\overline{S_0}-0.5\,.
\end{equation}
By differentiating this expression we get the following relationship for noise on the phase due to small signal fluctuations $\delta S$:
\begin{equation}
	\delta\varphi = \delta S/\overline{S_0}.
\end{equation}
Different noise sources can be considered, in particular detector readout noise ($\delta S_D$) and photon noise ($\delta S_P^2=S = 0.5\overline{S_0}$ when $\varphi=0$).
Denoting $\sigma^2$ the variance of the phase across the pupil and denoting the different contributors by the subscripts used above, we can write the following error budget for detection noise:
\begin{equation}
\sigma_D^2 + \sigma_P^2 = \delta S_D^2/\overline{S_0}^2 + 0.5/\overline{S_0}\,.
\end{equation}
In terms of noise propagation, our concept proves to be the most optimal wavefront sensor among the existing systems as shown by \citet{2005ApJ...629..592G}.

%%%%%%%%%%%%%%%%%%%%%%%%%%%%%%%%%%%%%%%%%%%%%%%%%%
\section{Static phase map reconstruction}\label{sec:reconstruction}
\begin{figure}[!ht]
\centering
\resizebox{\hsize}{!}{
\includegraphics{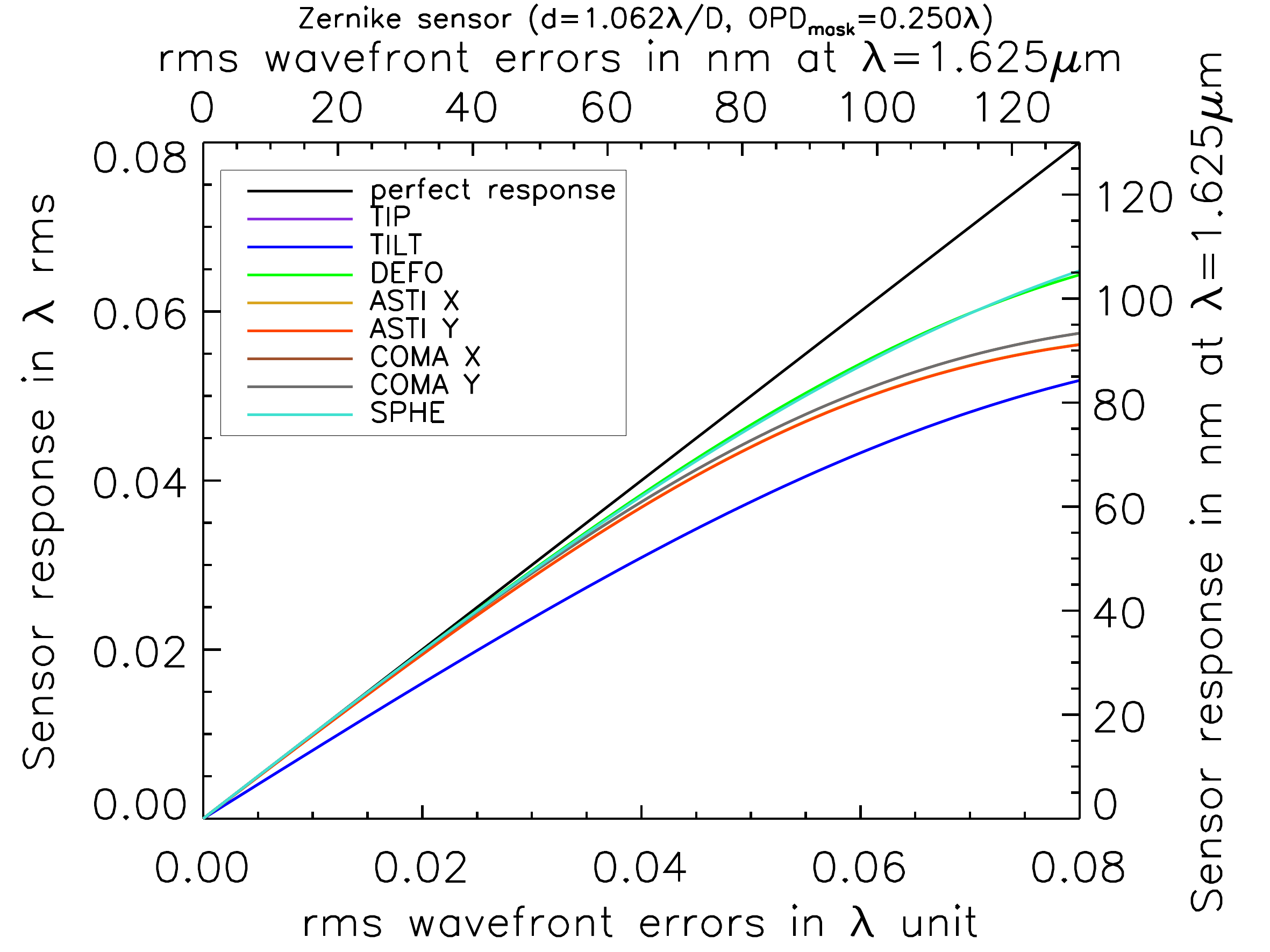}
}
\caption{Response of the Zernike sensor to residual wavefront errors for different low-order aberrations.} 
\label{fig:zernike_response}
\end{figure}

To illustrate wavefront reconstruction with the Zernike sensor by numerical simulations, we first consider its response to low-order aberrations. We apply increasing amounts of each of the first eight Zernike coefficients to an otherwise flat wavefront,  see Fig. \ref{fig:zernike_response}. For all the considered modes, a linear response is observed for values lower than 48\,nm (0.03\,$\lambda$) rms. Within this range and except for the tip/tilt aberrations, the curve of the Zernike sensor response to low-order aberrations is linear with a slope of unity. For tip/tilt aberrations (lateral image movement), a slope of 0.81 is observed. This curious exception is believed to be due to the modification of the light distribution going through the decentered mask. Carefully calibrated, this effect will have no consequence upon the capacity of the device to measure tip-tilt errors.

%_____________________________________________________________
%_____________________________________________________________

We then create an arbitrary phase map with a standard deviation $\sigma_{OPD}=$44\,nm (0.027\,$\lambda$) rms and peak-to-valley $PtV=$305\,nm (0.2\,$\lambda$) at $\lambda=$1.625\,$\mu$m. A $\pi/2$ phase mask with a diameter of 1.06\,$\lambda/D$ is applied. 

\begin{figure*}[!ht]
\centering
\resizebox{\hsize}{!}{
\includegraphics[width=12cm]{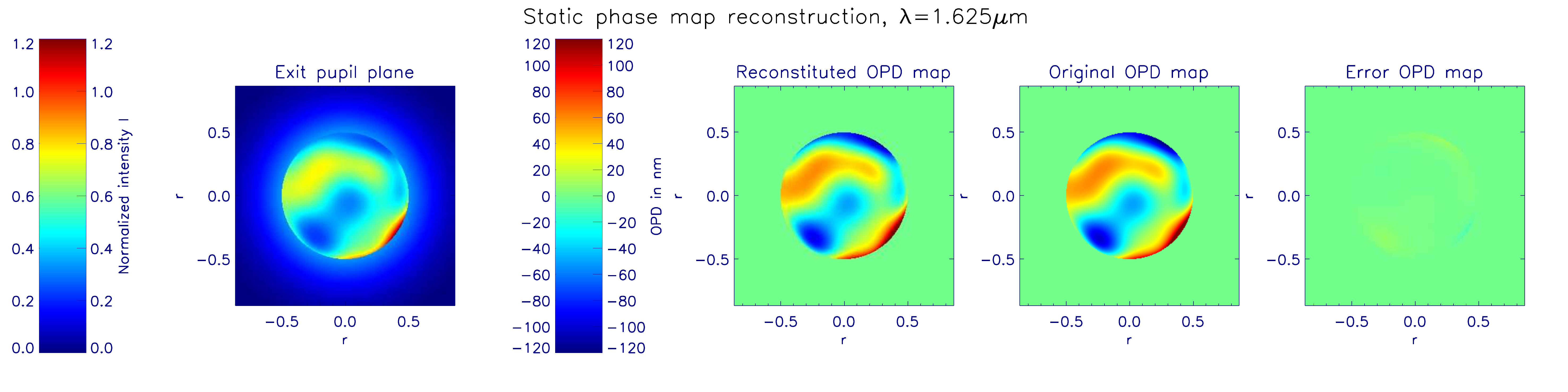}
}
\caption{Illustration of the concept of the Zernike phase mask sensor. From the left to the right: image in the exit pupil plane, the reconstituted OPD map, the original OPD map and the error OPD map.} 
\label{fig:imagesplaneC}
\end{figure*}

Figure \ref{fig:imagesplaneC} shows four panels from the left representing the pupil intensity map, the reconstituted phase map, the original phase screen, and the error map. The reconstituted map in the second panel is derived from the intensity map using the second-order formalism described above. The resulting phase map compares very well with the original, its standard deviation of $\sigma_{OPD}=$44\,nm rms is fully consistent with the value of the initial phase screen. The error map in the fourth panel, representing the difference between the original and reconstituted phase maps, gives a null mean value and an error of 1.9\,nm rms (0.0012\,$\lambda$), underlining the nanometric accuracy of the reconstitution with our concept. In addition, we underline the quasi-linearity of the wavefront error measurements with the Zernike sensor by noting the similarity of the pupil intensity map and the phase maps.
%_____________________________________________________________
%_____________________________________________________________

\begin{figure}[!ht]
\centering
\resizebox{\hsize}{!}{
\includegraphics{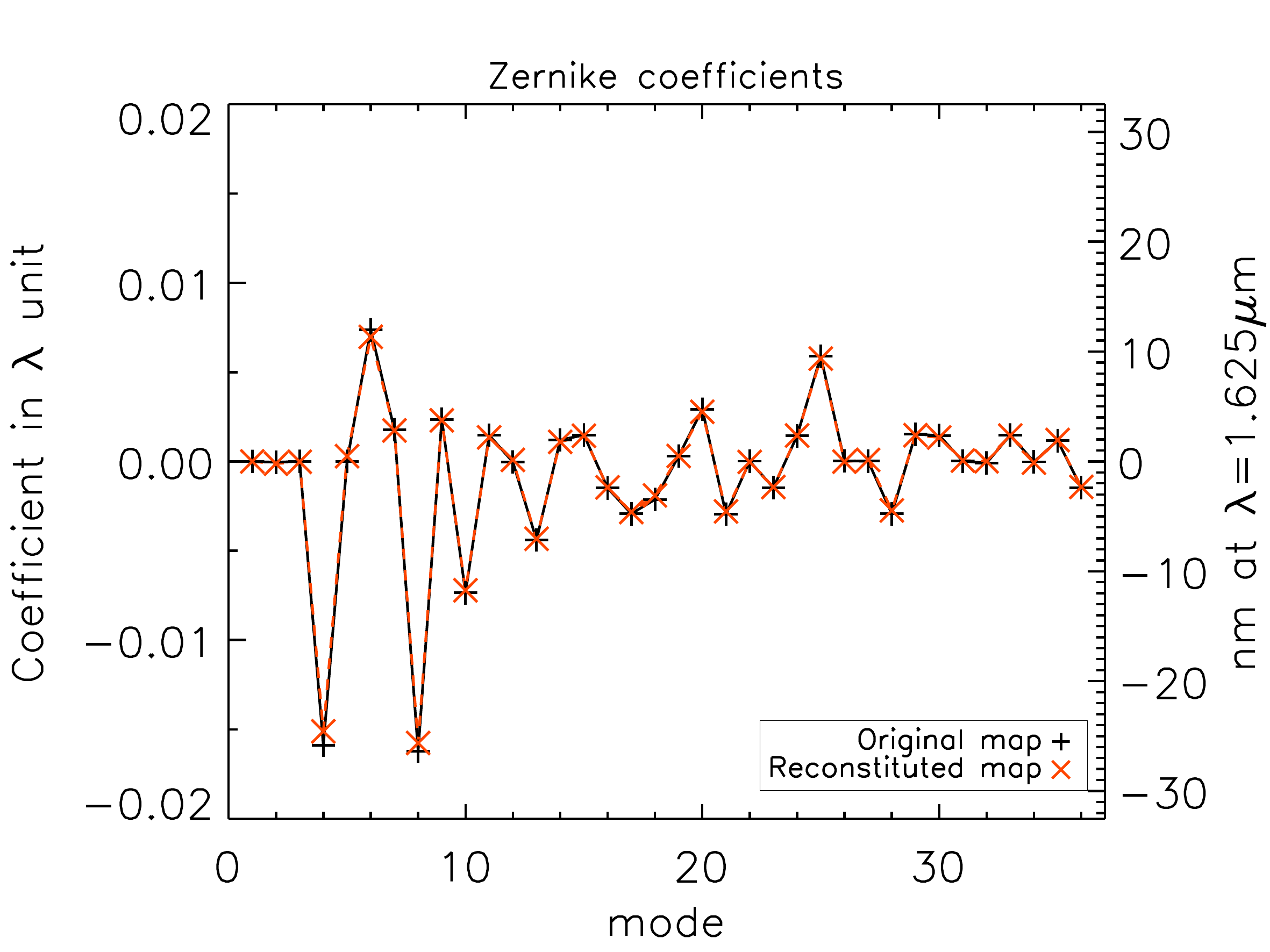}
}
\caption{Zernike coefficients of the original and reconstituted maps as a function of the aberration modes. The coefficients are ordered following the convention in \citet{1992ost..book.....M}.} 
\label{fig:zernike_coeffs}
\end{figure}

The excellent reconstitution of the wavefront can also be observed in Fig. \ref{fig:zernike_coeffs} where we plot the value of the first 36 Zernike aberration coefficients for the initial and estimated phase maps. 

%_____________________________________________________________
%_____________________________________________________________

%%%%%%%%%%%%%%%%%%%%%%%%%%%%%%%%%%%%%%%%%%%%%%%%%%
\section{Phase map reconstruction in the presence of AO residual}\label{sec:AOresidual}
\begin{figure*}[!ht]
\centering
\resizebox{\hsize}{!}{
\includegraphics{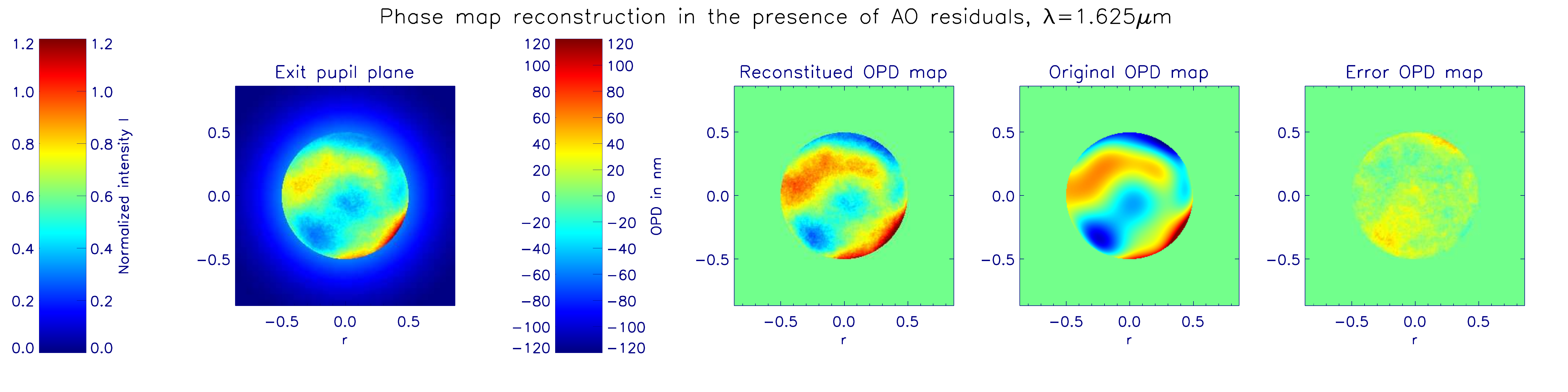}
}
\caption{Same as Fig. \ref{fig:imagesplaneC} but in the presence of 100 phase screens with 0.05\,$\lambda$ rms wavefront errors in the entrance pupil.} 
\label{fig:zernike_random_aberrations}
\end{figure*}

A typical application of this wavefront sensor is to measure static or quasi-static aberrations in the presence of rapidly varying aberrations. In particular, we want to calibrate the NCPA of an XAO system during observations. In the following we consider how the residual wavefront errors of the XAO correction affects such measurements. Keeping the phase map used previously as a static wavefront error and adding N independent phase screens sequentially allows us to determine the effect of atmospheric residuals on long exposure measurements. For simplicity, we assume AO residuals with $\nu^{-2}$-power spectral density (PSD) distributions, where $\nu$ denotes the spatial frequency of the aberration within the pupil. This is a good approximation of the spatial behavior of the AO-corrected phase screens up to the AO cut-off frequency ($1/2p$, $p$ denoting the inter-actuator pitch of the deformable mirror [DM]), which is the range of frequencies of interest for our application.
Figure \ref{fig:zernike_random_aberrations} illustrates the reconstitution of the quasi-static phase map in the presence of 100 phase screens with 81\,nm (0.05\,$\lambda$) rms wavefront error. The error map shows a standard deviation of 9.8\,nm ($6\times 10^{-3}\lambda$) rms. In addition to the expected random error map, we clearly see a deterministic component correlated with the original phase map. This reminiscence is due to the nonlinearity of the response curve, leading to an imperfect averaging of the AO residual phase.

%_____________________________________________________________
%_____________________________________________________________

\begin{figure}[!ht]
\centering
\resizebox{\hsize}{!}{
\includegraphics{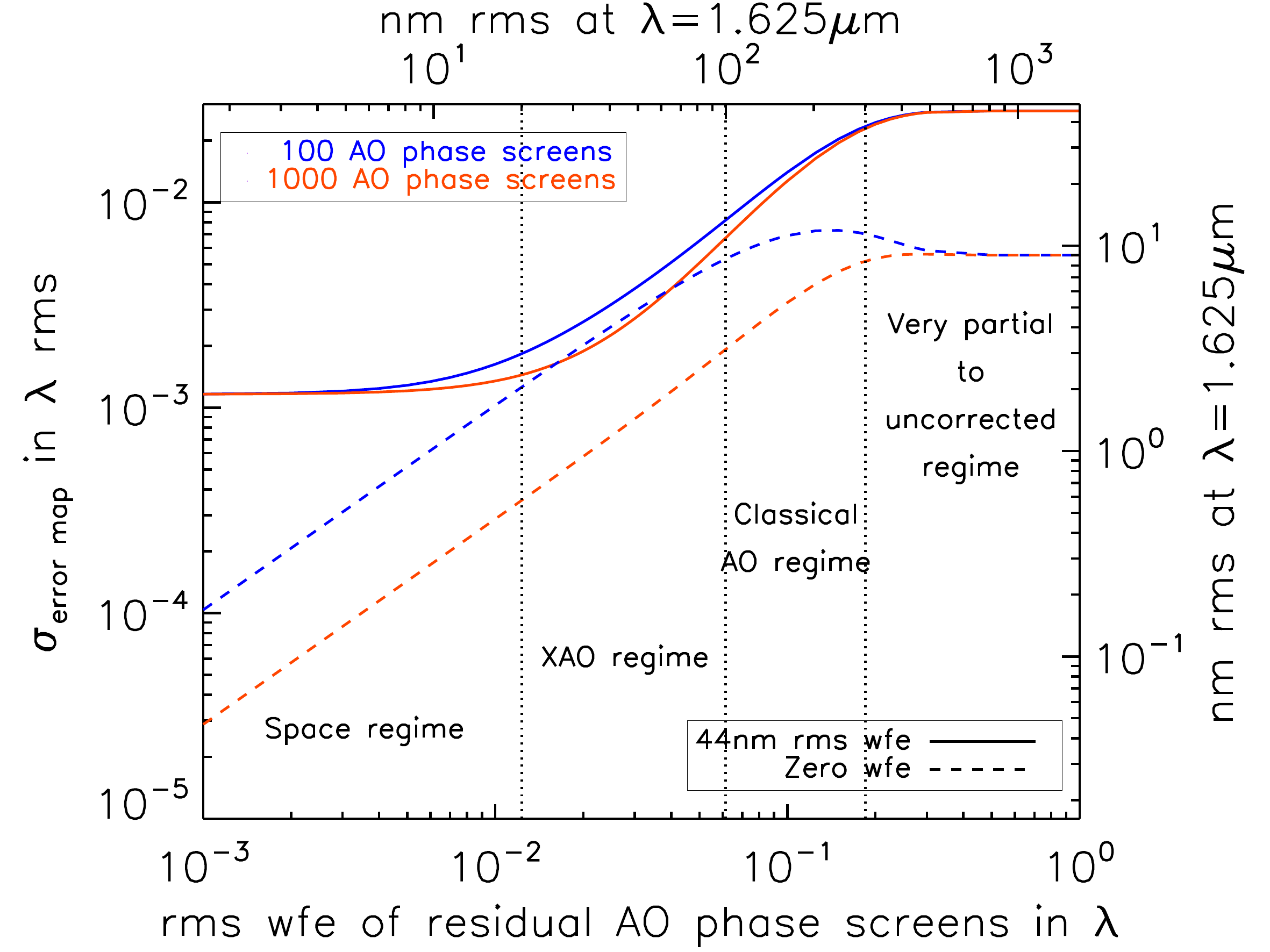}
}
\caption{Reconstitution error for a static phase map (\textit{solid line}) and zero NCPA (\textit{dashed line}) as a function of the amount of AO residuals. The dotted line corresponds to the static reconstitution error for the 44\,nm rms NCPA map.} 
\label{fig:zernike_random_aberrations_plot}
\end{figure}

Figure \ref{fig:zernike_random_aberrations_plot} shows the standard deviation of the error map as a function of the AO-residual wavefront error for 100 and 1000 phase screens (solid lines). Three ranges of AO residuals can be identified in this plot: a constant level for very small AO residuals corresponding to the static reconstruction error (1.9\,nm rms) reported above, and a nearly linear increase up to a saturation level corresponding to the aberration of the original static screen (44\,nm rms). In the intermediate regime, corresponding to a range of AO residuals between about 12\,nm and 240\,nm rms, the error is approximately equal to 1/10 of the amount of the AO residual independently of the number of phase screens. This level is close to what would be expected when averaging 100 phase screens (broken lines) but too high in the case of 1000 screens, probably due to the nonlinearity effect observed in Fig. \ref{fig:zernike_random_aberrations}.

In practical cases, the measurement made by the aid of this sensor will be fed back into the AO system in the form of offset voltages for the deformable mirror, leading to a near zero NCPA. Then the nonlinearity effect will be strongly reduced. To illuminate this case, Fig. \ref{fig:zernike_random_aberrations_plot} also plots the standard deviation of the error map as a function of the AO residual wavefront error for zero NCPA, showing a perfect fit with the expected $1/\sqrt{N}$ relationship (broken lines) up to about 240\,nm (0.15\,$\lambda$) rms. Assuming 90\,nm (0.055\,$\lambda_0$) AO residual aberrations for zero NCPA, an error of $1.7\times 10^{-3}\lambda_0$ is estimated for 1000 AO phase screens. Extrapolating this result for $N$ phase screens gives us a reconstruction error $\sigma=5.4\times 10^{-2}\lambda_0/\sqrt{N}$. The error for very large residuals, for which the wavefront sensor breaks down, is independent of the number of screens. 

The number of phase screens $N$ that are averaged during an exposure time $T$ is $N=T/\tau$, where $\tau$ is the lifetime of the phase screen. In current-generation XAO systems such as SPHERE, operating at around 1\,kHz, the total adaptive optics wavefront error budget amounts to some 55\,nm rms \citep{2006OExpr..14.7515F}, of which about $w_{SH} = 50$\,nm rms is attributed to sources such as Shack-Hartmann wavefront sensor noise propagation effects. The wavefront sensor noise is an uncorrelated noise related to the operating frequency $\nu_{XAO}$ of the XAO system. Considering $\nu_{XAO}=1$\,kHz, the aberrations due to the wavefront sensor noise has a lifetime $\tau_{SH}=1/\nu_{XAO}=1$\,ms. A 1\,s exposure therefore corresponds to 1000 noise error screens.

A part of the remaining AO errors, estimated to 20\,nm rms in the SPHERE budget \citep{2006OExpr..14.7515F}, is attributed to lag in the servo loop. As described in \citet{2005SPIE.5903..170M}, these errors create long-lived (order of 1s) speckles in coronagraphic images due to aberrations caused by time lag in classical AO systems as atmospheric phase screens blow across the telescope pupil. While advances in AO control technology, in particular the predictive Fourier control method \citep{2007JOSAA..24.2645P}, hold good promise of reducing the importance of such aberrations, this technique is not implemented in the SPHERE baseline. 

It has been pointed out to us that since this effect appears at a frequency similar to the one we propose for measuring NCPA using the Zernike sensor, this residual aberration will neither be corrected nor averaged out, hence constituting a limiting noise source. This is not the case, however, since although the speckles in the image plane have a long lifetime due to the translation of a quasi-static phase screen across the pupil, wavefront error observed by an individual detector pixel located in the pupil plane is not static. Indeed, while the corresponding speckle lifetime is on the order of $D/v$, with $D$ the telescope diameter and $v$ the wind speed, the life-time of an aberration observed in a pupil-plane pixel is $D/(vN_{pix})$, where $N_{pix}$ is the number of pixels across the pupil. In the case of SPHERE, assuming $N_{pix}=40$ for the Zernike sensor in order to match the spatial sampling of the XAO system, the pupil-plane lifetime of this effect is on the order of $\tau_{Lag} = $25\,ms.\\
Denoting aberration lifetime as $\tau$ and exposure time as $T$, the Zernike sensor measurement error variance due to an aberration $w$ is $w^2 \tau/T$. We therefore identify two main contributors of AO residual aberrations to the error budget:\\
\begin{equation}
	\sigma_{AO}^2=\sigma_{Lag}^2 + \sigma_{SH}^2 = (\tau_{Lag} w_{Lag}^2+\tau_{SH} w_{SH}^2)/T\,.
\end{equation}

%_____________________________________________________________
%_____________________________________________________________
 
%%%%%%%%%%%%%%%%%%%%%%%%%%%%%%%%%%%%%%%%%%%%%%%%%%
\section{Sensitivity to chromatic effects}\label{sec:chromaticity}
\begin{figure}[!ht]
\centering
\resizebox{\hsize}{!}{
\includegraphics{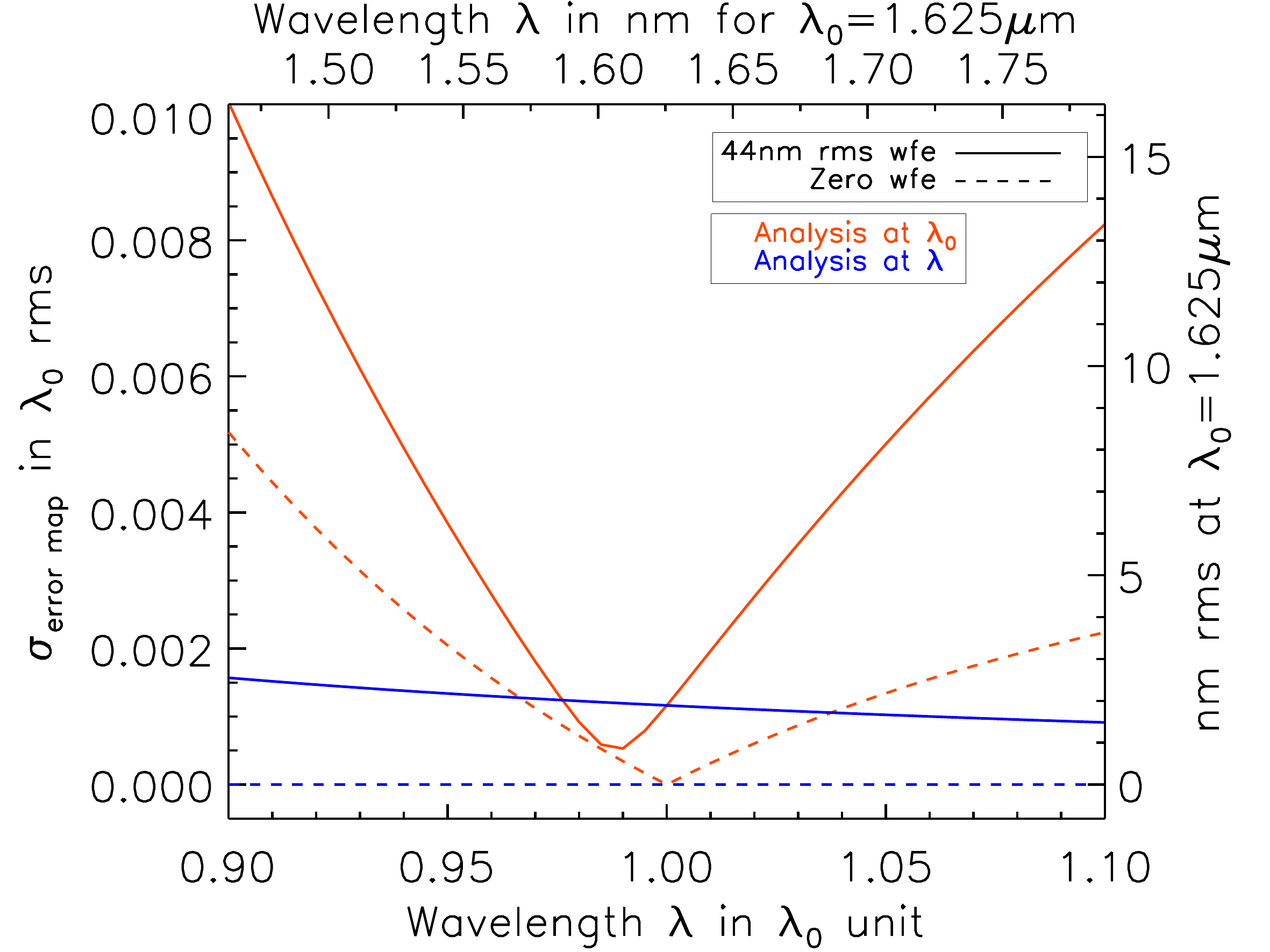}
}
\caption{Wavefront error measurement as a function of the source wavelength $\lambda$ for an analysis at $\lambda_0$.} 
\label{fig:zernike_chromaticity_nosum_plot}
\end{figure}

The above results of the Zernike sensor have been given for a monochromatic light source emitting at wavelength $\lambda_0$, which has been used for the definition of the phase mask and for the data analysis. We now study the impact of chromatic effects.
 We first study the impact of a source wavelength $\lambda \neq \lambda_0$ on the Zernike sensor measurement for an analysis done at $\lambda_0$, see Fig. \ref{fig:zernike_chromaticity_nosum_plot}. In the case of zero aberration, the measurement error is minimal at $\lambda_0$ and increases quickly when the analysis wavelength moves away from the design wavelength. However, if we know the source wavelength precisely and recalculate the $\theta$ and $b$ terms accordingly, the residual error is reduced to the numerical noise level for the entire wavelength range considered here.\\ 
When the nonzero aberration case is considered, as shown by the red solid-line curve in Fig. \ref{fig:zernike_chromaticity_nosum_plot}, a steeper rise from the optimal wavelength is observed, but, more remarkably, the optimal wavelength is blue-shifted by about 1\%. This can be explained by the fact that the reduced Strehl ratio is partly compensated for by an increased concentration of flux within the Zernike mask. Assuming a realistic situation where the analysis wavelength is known and taken into account for the zero-aberration case, the measurement errors in the presence of $0.027\lambda_0$ static aberrations are smaller than $1.5\times 10^{-3}\lambda_0$ within the range considered, see Fig. \ref{fig:zernike_chromaticity_nosum_plot}. Possible improvements in this value can be obtained, working at an analysis wavelength $\sim$1\% shorter than the source wavelength as shown in the plot where the red-line curve line is found below the blue-line one within the range between 0.98$\,\lambda_0$ and $\lambda_0$, offering possible advanced strategies of analysis, but this goes beyond the scope of this paper.
%_____________________________________________________________
%_____________________________________________________________

\begin{figure}[!ht]
\centering
\resizebox{\hsize}{!}{
\includegraphics{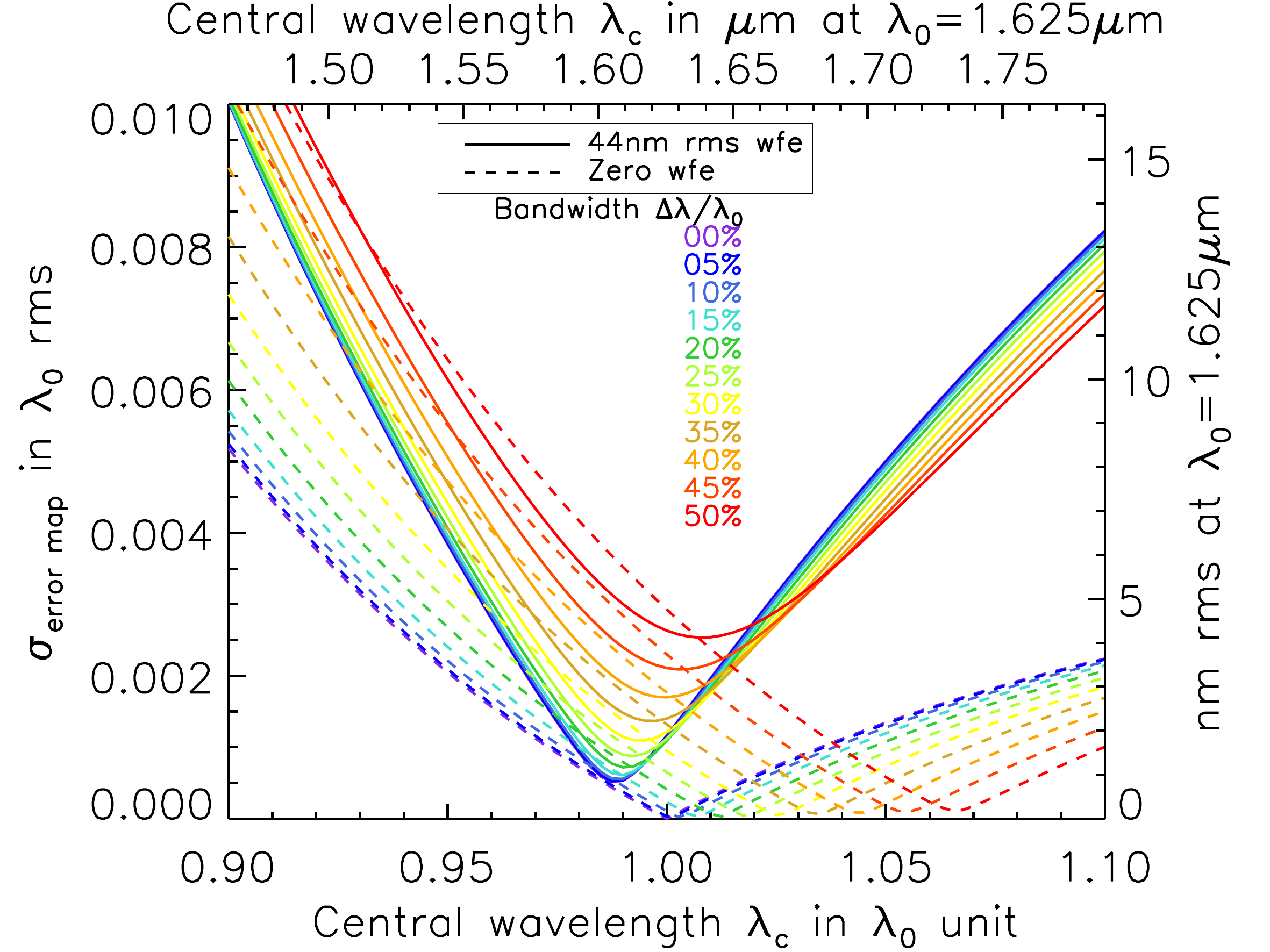}
}
\caption{Wavefront error measurement as a function of the central wavelength $\lambda_c$ for different spectral bandwidths.} 
\label{fig:zernike_chromaticity_zero_map_lambdac_plot}
\end{figure}

\begin{figure}[!ht]
\centering
\resizebox{\hsize}{!}{
\includegraphics{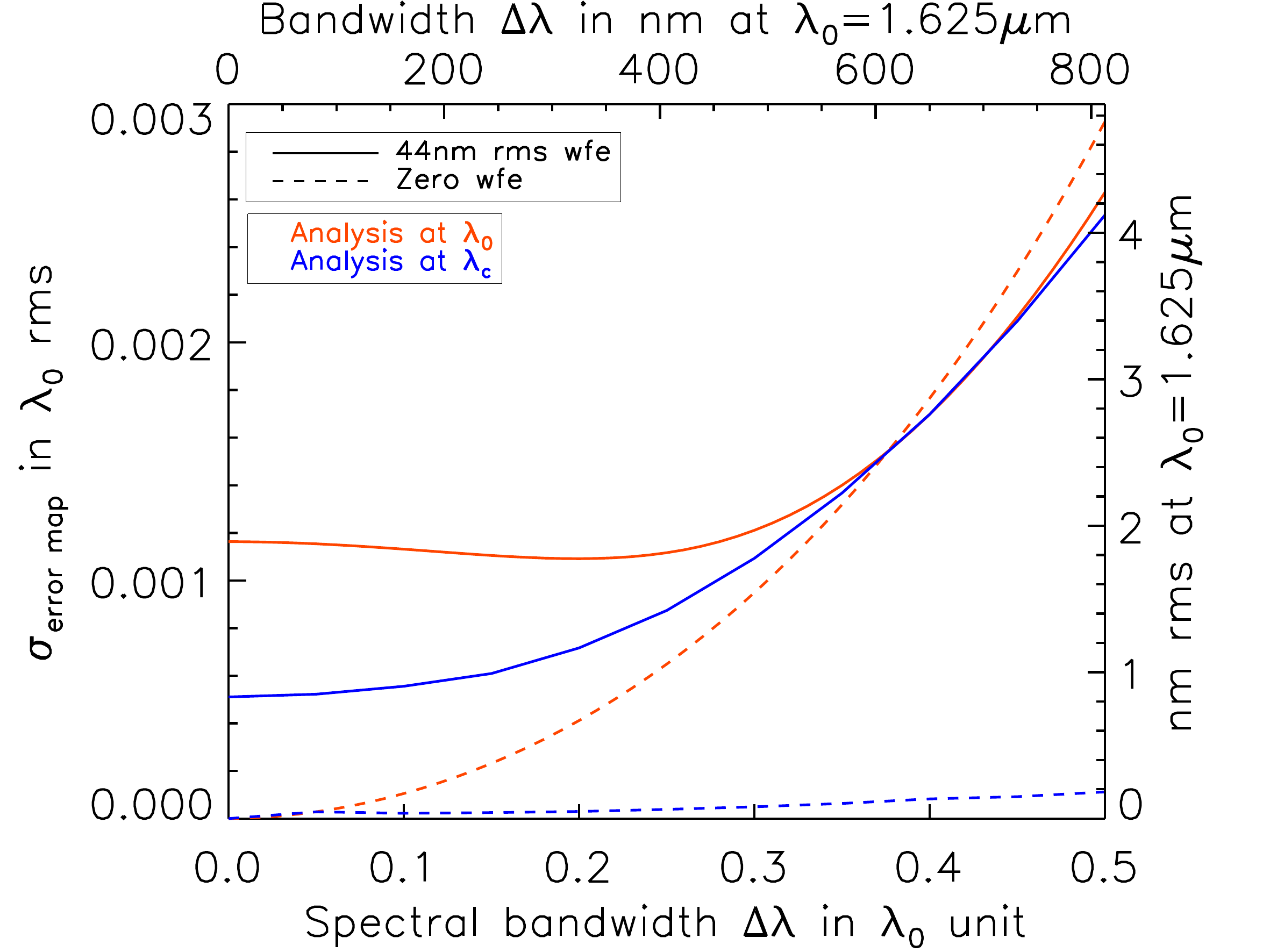}
}
\caption{Wavefront error measurement as a function of the spectral bandwidth $\Delta\lambda$ for an analysis at $\lambda_0$ and at the optimal central wavelength $\lambda_c$.}
\label{fig:zernike_chromaticity_poly_plot}
\end{figure}

In broadband, assuming a flat spectral distribution over a spectal range $\Delta\lambda$ centred at $\lambda_c$, the zero-aberration error has the same shape as in the monochromatic case, but it is red shifted, see Fig. \ref{fig:zernike_chromaticity_zero_map_lambdac_plot}. The minimum remains very low, reaching $2\times 10^{-4}\lambda_0$ for 50\% bandwidth as seen in Fig. \ref{fig:zernike_chromaticity_poly_plot}. In this case the red shift is 6.5\%. In the presence of aberrations, the shift is smaller, compensating for the monochromatic blueshift, but the increase in measurement error is larger. We attribute these effects to the shape of the monochromatic curves observed in Fig. \ref{fig:zernike_chromaticity_nosum_plot}: the bandwidth-induced shift is caused by the asymmetry of these curves, more pronounced in the unaberrated case, and the measurement error is related to their steepness, more pronounced in the aberrated case.\\
Assuming a realistic case of 20\% bandwidth, corresponding roughly to the atmospheric H-band centered on 1625\,nm, and using the central wavelength as the analysis wavelength, the chromatic measurement error for zero aberrations is $\sigma_{\lambda} = 4.1\times 10^{-4}\lambda_0$, corresponding to 0.65\,nm, when using the central wavelength as the analysis wavelength, but can be reduced to $3.0\times 10^{-5}\lambda_0$ if the filter function and object spectral type are known, see Fig. \ref{fig:zernike_chromaticity_poly_plot}. As discussed in the introduction, we must account for the calibration of a small differential aberration between the Zernike sensor plane and the coronagraph plane of 1 to 2\,nm rms. This is small enough not to have a significant impact on this budget, as can be estimated from the curves in Fig. \ref{fig:zernike_chromaticity_poly_plot}. We therefore assume the value of $\sigma_{\lambda}=3.0\times 10^{-5}\lambda_0$ for the error budget explained below.
  
%_____________________________________________________________
%_____________________________________________________________

%_____________________________________________________________
%_____________________________________________________________

%%%%%%%%%%%%%%%%%%%%%%%%%%%%%%%%%%%%%%%%%%%%%%%%%%
\section{Application to a real system, error budget}\label{sec:real_system}
We here consider the application of this sensor in a typical XAO system similar to the ones currently being constructed \citep{2008SPIE.7015E..31M,2010SPIE.7736E..13S}. These are optimized to work in the near infrared around $\lambda = 1.6\,\mu$m with atmospheric correction operating at frequencies exceeding 1\,kHz. NCPA are calibrated off-line at intervals limited by practical and observing efficiency considerations to more than one hour, perhaps as much as one day. Even though the instrument is expected to be quite stable during this time frame, evolutions in NCPA due to thermo-elastic effects, rotating elements such as the atmospheric dispersion correctors, and air-mass related chromatic beam shift are inevitable. For the purpose of SPHERE error budgets \citep{KD..victoria}, such evolutions have been estimated to around 15\,nm rms, although actual performance of the as-built system is expected to be better than this.
We investigate the accuracy of the Zernike sensor introduced in an exoplanet direct-imaging instrument, considering several error sources. We assume a high-quality beamsplitter, polished to nanometric surface accuracy, separating the measurement beam from the science beam just upstream of the coronagraph mask. When aberrations are measured during observations and fed back to the XAO system in the form of updated reference slopes, the aberrations measured by the sensor will essentially be zero so that we can ignore error terms related to the absolute level of wavefront distortion. To this end, an error budget is built in the context of a SPHERE upgrade based on the terms identified throughout this paper:

%_____________________________________________________________
\begin{equation}
\begin{split}
	\sigma^2  = & \sigma_D^2 + \sigma_P^2 + \sigma_{AO}^2 + \sigma_\lambda^2\\ 
					  = &  \delta S_D^2/\overline{S_0}^2 +0.5/\overline{S_0} \\
					 & + (\tau_{Lag} w_{Lag}^2+\tau_{SH} w_{SH}^2)/T + (3.0\times 10^{-5}\lambda_0)^2\,.
\end{split}
\label{eq:error_budget}
\end{equation}
%_____________________________________________________________

\begin{figure}[!ht]
\centering
\resizebox{\hsize}{!}{
\includegraphics{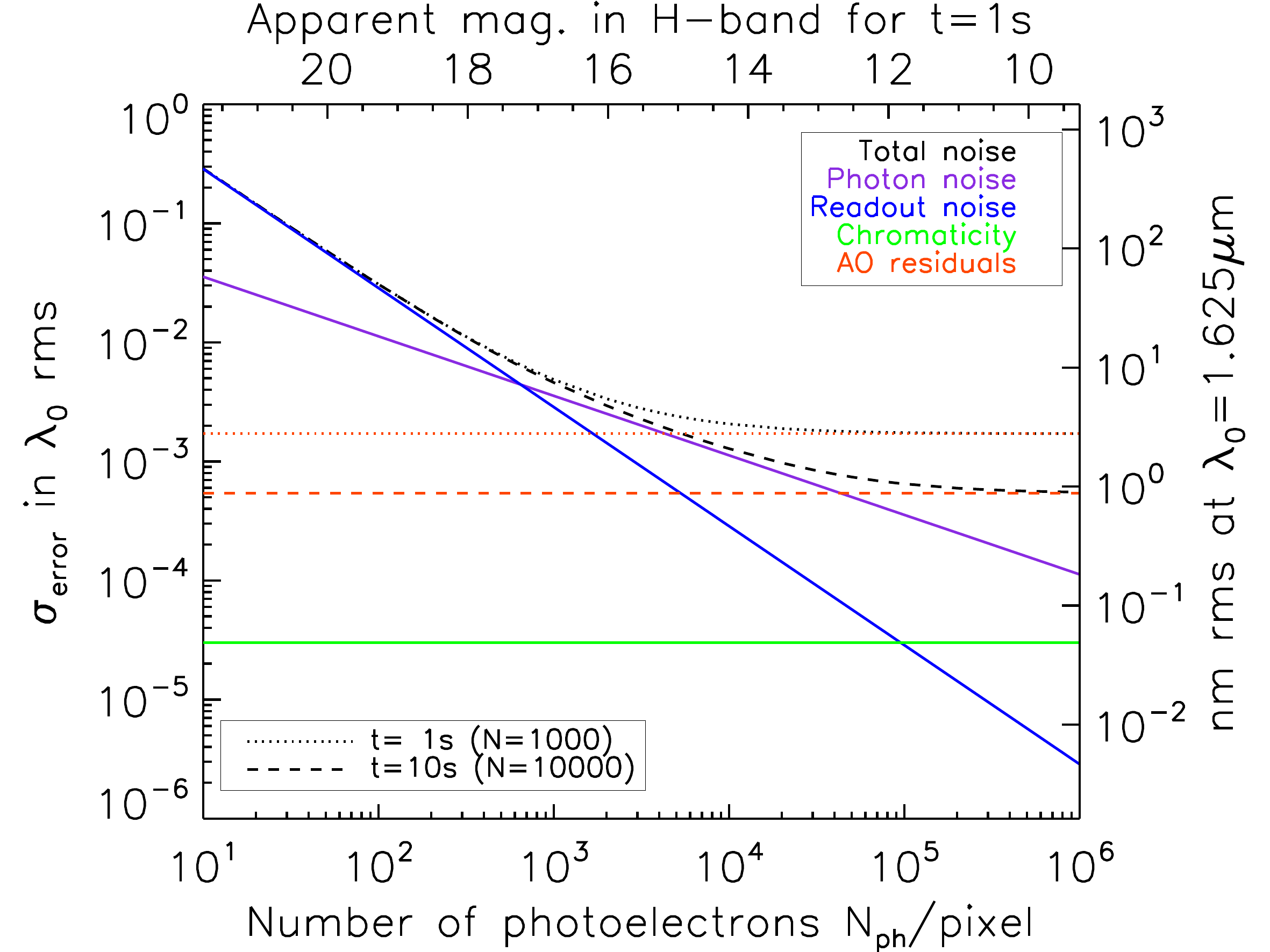}
}
\caption{Wavefront error measurement in the presence of different noise sources as a function of the number of photo-electrons for a 40$\times$40 pixel pupil size.} 
\label{fig:error_budget_plot}
\end{figure}

The different parameters for our calculation are listed in Table \ref{table:parameters} while the different contributors to the measurement errors for our concept are reported in Table \ref{table:error_budget}. The AO cut-off frequency of SPHERE is $\nu_C =1/2p$ with p=20cm, and so a pupil of 40$\times$40 pixels (or subpupils) is required for the Zernike sensor to perform measurement up to this frequency. We consider two different exposure times in this study, 1\,s and 10\,s.\\
Figure \ref{fig:error_budget_plot} displays the resulting measurement error of the NCPA map as a function of number of photo-electrons per pixel. At low flux, the measurement error is dominated by readout noise. At intermediate flux range, all the noise sources contribute equivalently, leading to a total noise of 0.01\,$\lambda$ (16\,nm) rms in the presence of 600\,e$^{-}$ per pixel. At high flux, the error measurement reaches a plateau of $2\times 10^{-3}\,\lambda$ (3\,nm) rms for 1\,s exposure and $5\times 10^{-4}\,\lambda$ (1\,nm) for 10\,s exposure, representing the time-lag part of the AO residuals that dominates the noise in this range. It should be noted that this limit is set by the XAO system rather than the Zernike sensor and that, as mentioned above, techniques such as predictive Fourier control method for AO control are expected to efficiently reduce such AO residuals \citep{2007JOSAA..24.2645P}. Implementation of this concept in SPHERE as an upgrade path 
appears as an efficient and realistic means of obtaining subnanometric precision for NCPA compensation in the H-band, representing an improvement in terms of wavefront error by a factor of 3 to 10. According to \citet{2003ApJ...596..702P}, the residual speckles in a highly corrected coronagraphic image are proportional to the power spectral density of the wavefront, hence proportional to the square of the rms wfe. We can therefore expect a reduction of the residual speckles in SPHERE images by a factor ranging from 10 to 100. 

%_____________________________________________________________
\begin{table}
\caption{Parameters used for estimating the error budget.}
%\centering
\begin{tabular}{l l l}
\hline\hline
\textbf{Parameters} & \textbf{Relative value} & \textbf{Physical value}\\
\hline
Zero mag. flux density in H-band & & 1080\,Jy\\
Telescope diameter $D$ & & 8\,m\\
Telescope transmission $T_{tel}$ & 40\% & \\
Beamsplitter transmission $T_{BS}$ & 5\% & \\
Mask design wavelength $\lambda_0$ & & 1.625\,$\mu$m\\
Focal ratio F/\#  & & 40\\
Mask diameter $d$ & 1.062\,$F\lambda_0$ & 69.0\,$\mu$m\\
Mask phase shift $\theta$ & $\pi/2$ & \\
Refractive index $n$ (Fused silica)& & 1.4431 @ $\lambda_0$\\
Mask depth z & 0.25\,$\lambda_0$/(n-1) & 0.917\,$\mu$m\\
Filter wavelength $\lambda_c$	 & $\lambda_0$ & 1.625\,$\mu$m\\
Filter bandwidth $\Delta\lambda$ & 20$\%$ & 0.325\,$\mu$m\\
DM inter-actuator pitch $p$ & & 20cm\\  
AO residual amount & 0.055\,$\lambda_0$ & 90\,nm\\
Wind speed $v$ & & 10\,m.s$^{-1}$\\
HAWAII-2RG readout noise & & 18\,e$^{-}$\\
HAWAII-2RG pixel full well & & $10^{5}$\,e$^{-}$\\
Pupil size & & $40\times 40$ pixels\\
\hline
\end{tabular}\\
\label{table:parameters}
\end{table}
%_____________________________________________________________

%_____________________________________________________________
\begin{table}
\caption{Error budget. The term $\overline{S_0}$ denotes the average entrance pupil flux.}
%\centering
\begin{tabular}{l l}
\hline\hline
\textbf{Error} & \textbf{Value in e$^{-}$}\\
\hline
\tablefootmark{a}Photon noise $\sigma_P$ & $\sqrt{0.5\,\overline{S_0}}$\\
\tablefootmark{a,b}Readout noise $\sigma_D$ & 18\,e$^{-}$\\
\tablefootmark{c}AO residuals error $\sigma_{AO}$ & $2\pi\times 5.4\times 10^{-2}\overline{S_0}/\sqrt{N}$\\
\tablefootmark{d}Chromaticity $\sigma_\lambda$ & $2\pi\times 3.0\times 10^{-5}\overline{S_0}$\\
\hline
\end{tabular}\\
\tablefoottext{a}{From Sect. \ref{subsec:noise_propagation}.}\\
\tablefoottext{b}{From Table \ref{table:parameters}.}\\
\tablefoottext{c}{From Sect. \ref{sec:AOresidual}.}\\
\tablefoottext{d}{From Sect. \ref{sec:chromaticity}.}\\
\label{table:error_budget}
\end{table}
%_____________________________________________________________

%_____________________________________________________________
%_____________________________________________________________

%%%%%%%%%%%%%%%%%%%%%%%%%%%%%%%%%%%%%%%%%%%%%%%%%%%%%%%%%%%%%
\section{Conclusion}\label{sec:Conclusion}
In an exoplanet imager, non-common path aberrations between the visible XAO sensing path and the near-infrared scientific path induce quasi-static speckles in a coronagraphic image, making the observation of faint exoplanets impossible. An excellent calibration of these quasi-static aberrations due to chromatic differential optics is required to correct these residual speckles. For this purpose, we have developed the Zernike phase-mask sensor, revisiting the phase-contrast method of \citet{1934MNRAS..94..377Z} in the context of exoplanet direct imaging. This system uses a focal plane phase mask operating in the same wavelength as the coronagraph to encode the NCPA present in the upstream pupil plane into intensity variations in the relayed pupil. 

We established a formalism for this approach, considering the spatial variability of the wave diffracted by the mask. The importance of this parameter, often ignored or simplified in classical textbooks \citep{hecht1987optics,1992ost..book.....M,1996ifo..book.....G,1999po..book.....B} and by contemporary authors \citep{2011SPIE.8126E..11W}, is underlined here, allowing us to reach an accurate calibration of residual wavefront errors with the Zernike sensor. A quasi-linear relationship between the NCPA map and the intensity variations in the exit pupil is obtained with this approach, providing a simple algorithm for the reconstitution of a static phase map. 

We analyzed the accuracy of the Zernike sensor for calibrating static aberrations and demonstrated its efficiency for online measurements of static or quasi-static aberrations in a typical XAO system. In the presence of AO residuals of 81\,nm, representative of the SPHERE system, static aberrations are reduced to around 1\,nm at $\lambda=1.6\,\mu$m for 10\,s integrations, leading to an attenuation of the residual speckles by a factor ranging from 10 to 100, compared with baseline SPHERE performance. 

Following this study, we are now preparing an experimental validation of the Zernike phase mask sensor. A $\pi/2$ phase mask has been manufactured, following the method described in \citet{2010A&A...509A...8N} for the Roddier \& Roddier phase mask. We are currently testing this prototype on our high-contrast imaging testbed at Marseilles, while studying the sensitivity of the concept to noise source, and the results will be presented in a forthcoming paper. 

The simplicity of the Zernike sensor design makes its implementation possible as an upgrade path for the forthcoming exoplanet imagers on the ground \citep[e.g. SPHERE, GPI, SCExAO, P1640,][]{2008SPIE.7014E..41B,2008SPIE.7015E..31M,2010SPIE.7736E..71G,2011PASP..123...74H}. We have recently inserted a dedicated prototype in SPHERE, allowing real-life tests during the instrument's integration phase. Its application also appears very interesting for future coronagraphic missions in space \citep{2010SPIE.7731E..68G,2010SPIE.7731E..67T,2012ExA...tmp...11B}, as well as for the next-generation exoplanet imager EPICS for the E-ELT \citep{2008SPIE.7015E..46K}. Its ability to work in the presence of a telescope aperture with segmented primary mirror, central obstruction, and spider arms makes the technique particularly promising for complex aperture geometries.

%%%%%%%%%%%%%%%%%%%%%%%%%%%%%%%%%%%%%%%%%%%%%%%%%%%%%%%%%%%%%
\begin{acknowledgements} 
This research was supported by funding from CNRS-INSU and the Institute Carnot STAR. The authors would like to thank the anonymous referee for the careful reading, suggestions, and comments on the manuscript, Jean-Fran\c{c}ois Sauvage and Laurent Mugnier for fruitful discussions as well as Marc Ferrari for his support, the R\'egion Provence-Alpes-C\^ote d'Azur and ONERA for financial support with B. Paul's scholarship. MN acknowledges Laurent Pueyo for his insightful commentaries on the manuscript and R\'emi Soummer for his support. This work is partially based upon work supported by the National Aeronautics and Space Administration under Grant NNX12AG05G issued through the Astrophysics Research and Analysis (APRA) program.
\end{acknowledgements}

%%%%%%%%%%%%%%%%%%%%%%%%%%%%%%%%%%%%%%%%%%%%%%%%%%%%%%%%%%%%%
%%%%% References %%%%%X
\bibliographystyle{aa}   %>>>> makes bibtex use spiebib.bst
\bibliography{articulo06_v29astro-ph}   %>>>> bibliography data in report.bib

\end{document}